\documentclass[12pt, centerh1]{article}
\usepackage{graphicx}

\usepackage{geometry,multirow}
\usepackage{graphicx}
\usepackage{epstopdf}
\usepackage{natbib}
\usepackage{bm}
\usepackage[latin1]{inputenc}
\usepackage{amssymb}
\usepackage{amsfonts, amsmath, amssymb, marvosym,colonequals}
\usepackage{algorithm}
\usepackage{bm}
\usepackage{mathtools}
\usepackage{caption}
\usepackage{theorem}
\usepackage{caption}

\newcommand{\tr}{\,\mbox{tr}}

\newcommand{\load}{\bm\Lambda}
\newcommand{\noisev}{\bm\Psi}

\newcommand{\ident}{\mathbf{I}}

\newcommand{\vecx}{\mathbf{x}}

\newcommand{\vecX}{\mathbf{X}}
\newcommand{\vecU}{\mathbf{U}}
\newcommand{\vecu}{\mathbf{u}}

\newcommand{\vecpi}{\bm\pi}
\newcommand{\vectheta}{\bm\theta}

\newcommand{\varthet}{\bm\vartheta}
\newcommand{\vecmu}{\bm\mu}
\newcommand{\vecepsilon}{\bm\epsilon}
\newcommand{\vecxi}{\bm\xi}

\newcommand{\vecalpha}{\bm\alpha}

\newcommand{\matsig}{\bm\Sigma}

\newcommand{\e}[1]{{\mathbb E} #1 }
\newcommand{\prev}{\mbox{\tiny prev}}

\begin{document}

\title{A Mixture of Generalized Hyperbolic Factor Analyzers}

\author{Cristina~Tortora,\thanks{Department of Mathematics \& Statistics, McMaster University, Hamilton, Ontario,  L8S 4L8, Canada. E-mail: ctortora@mcmaster.ca.}
      \  ~Paul~D.~McNicholas and Ryan~P.~Browne
}
\date{Department of Mathematics \& Statistics, McMaster University.}

\maketitle

\begin{abstract}
The mixture of factor analyzers model, which has been used successfully for the model-based clustering of high-dimensional data, is extended to generalized hyperbolic mixtures. The development of a mixture of generalized hyperbolic factor analyzers is outlined, drawing upon the relationship with the generalized inverse Gaussian distribution. An alternating expectation-conditional maximization algorithm is used for parameter estimation, and the Bayesian information criterion is used to select the number of factors as well as the number of components. The performance of our generalized hyperbolic factor analyzers model is illustrated on real and simulated data, where it performs favourably compared to its Gaussian analogue and other approaches.
\end{abstract}

\section{Introduction}

Finite mixture models assume that a population is a convex combination of a finite number of densities; therefore, they are a natural choice for classification and clustering applications. A random vector $\mathbf{X}$ follows a (parametric) finite mixture distribution if its density can be written 
$f(\vecx\mid\varthet)= \sum_{g=1}^G \pi_g f_g(\vecx~|~\vectheta_g),$ where $\pi_g >0$, such that $\sum_{g=1}^G \pi_g = 1$, is the $g$th mixing proportion, $f_g(\vecx\mid\vectheta_g)$ is the $g$th component density, and $\varthet=(\vecpi,\vectheta_1,\ldots,\vectheta_G)$ is the vector of parameters, with $\vecpi=(\pi_1,\ldots,\pi_G)$. The component densities $f_1(\vecx~|~\vectheta_1),\ldots,f_G(\vecx\mid\vectheta_G)$ are usually taken to be of the same type and, until quite recently, the Gaussian mixture model has dominated the model-based clustering and classification literature  \cite[e.g.,][]{mclachlan00a,McLachlan03,Bouveyron07,McNicholas08,McNicholas10,Baek10,Montanari11,bhattacharya14,browne14,wei14}. The density of a Gaussian mixture model is $f(\vecx\mid\varthet)= \sum_{g=1}^G \pi_g\phi(\vecx\mid\vecmu_g,\matsig_g),$ where $\phi(\vecx\mid\vecmu_g,\matsig_g)$ is the multivariate Gaussian density with mean~$\vecmu_g$ and covariance matrix $\matsig_g$. The use of mixture models for clustering is known as model-based clustering, and model-based classification is the semi-supervised analogue.

Over the past few years, non-Gaussian model-based clustering techniques have gained popularity because Gaussian mixtures do not necessarily yield satisfactory results when clusters are asymmetric and/or have longer tails. The vast majority of non-Gaussian model-based clustering work to date has taken place over the past few years \citep[e.g.,][]{karlis09,lin09,lin10,browne11,vrbik12,vrbik14,smcnicholas13,lee13,morris13b,morris13a,franczak14,murray14a,murray14b,subedi14,ohagan15}.
The first non-Gaussian analogue of the mixture of factor analyzers model \citep{ghahramani97} was an extension to multivariate $t$-mixtures \citep{mclachlan07}, and this work was subsequently built on, remaining within the $t$-mixture framework \citep{andrews11b,andrews11c,andrews12,steane12,lin14}. 
Very recently, the mixture of factor analyzers model has been extended to mixtures of skew-$t$ distributions \citep{murray13c,murray14a}, mixtures of skew-normal distributions \citep{lin13}, and mixtures of shifted asymmetric Laplace distributions \citep{franczak13}. In this paper, we outline a more general case, i.e., their extension to generalized hyperbolic mixtures. 

The generalized hyperbolic distribution has the advantage of being very flexible; in fact, with specific (or limiting) values of the parameters, it can lead to other well known distributions. 
It can detect clusters with non-elliptical form because it contains a skewness parameter; in addition, there are concentration and index parameters \citep[cf.][]{BroMcN15}. The flexibility of this model comes at a modest cost --- one additional parameter per component --- relative to the mixture of skew-t  factor analyzers of \cite{murray14a}. In addition to the skew-$t$ distribution, other distributions that have been used for model-based clustering can be obtained as a special or limiting case of the generalized hyperbolic distribution, e.g., the $t$ distribution, the multivariate normal-inverse Gaussian distribution \citep[cf.][]{karlis09}, the variance-gamma distribution \citep[cf.][]{smcnicholas13}, and the asymmetric Laplace distribution \citep[cf.][]{franczak14}. 

The remainder of this paper is laid out as follows. In Section~\ref{sec:background}, we describe the mixture of generalized hyperbolic distributions. In Section~\ref{sec:method}, we outline the extension of the mixture of factor analyzers model to generalized hyperbolic mixtures. Our approach is illustrated in Sections~\ref{sec:data} and~\ref{sec:simulation}. We conclude with a summary and suggestions for future work (Section~\ref{sec:conc}).

\section{Background}\label{sec:background}

The density of a $p$-dimensional random variable $\bf X$ from a generalized hyperbolic distribution is
\begin{equation}\label{eqn:hyp} 
 f_{\text{H}}({\bf x}~|~{\bm \vartheta})=\left[\frac{\chi+\delta  ({\bf x}, {\bm\mu}|{\matsig})}{\varphi+{\bm \alpha}' {\bm\matsig}^{-1}{\bm \alpha}} \right]^{\frac{\lambda-{p}/{2}}{2}} 
\frac{\left({\varphi}/{\chi}\right)^{\frac{\lambda}{2}}K_{\lambda-\frac{p}{2}}\Big(\sqrt {[\varphi+{\bm\alpha}'{\bm\matsig}^{-1}{\bm \alpha}][\chi+\delta({\bf x},{\bm\mu}|{\bm\matsig})]}\Big)}
 {(2\pi)^{\frac{p}{2}}|{\bm\matsig}|^{\frac{1}{2}}K_\lambda\big(\sqrt{\chi\varphi}\big)\exp{\big\{({\bm\mu}-{\bf x})'{\bm\matsig}^{-1}{\bm\alpha}\big\}}},
\end{equation}
where $p$ is the number of variables, $\delta ({\bf x}, \bm\mu~|~\matsig)=({\bf x}-{\bm \mu})'{\matsig}^{-1}({\bf x}-{\bm \mu})$ is the squared Mahalanobis distance between ${\bf x}$ and~${\bm  \mu}$, $K_{\lambda}$ is the modified Bessel function of the third kind with index $\lambda$, and ${\bm  \vartheta}$ denotes the vector of parameters. The parameters have the following interpretation: $\lambda$ is an index parameter, $\chi $ and $\varphi$ are concentration parameters, $\bm \alpha$ is a skewness parameter, $ \bm\mu$ is a location parameter, and $\bm\matsig$ is a scale matrix.

Let $Y\sim\text{GIG}(\psi, \chi,\lambda)$, where GIG indicates the generalized inverse Gaussian distribution \citep{good53,barndorff77,blaesild78,halgreen79,jorgensen82} with density given by
\begin{eqnarray}
\label{GIG}
h(y~|~\vecxi)=
\frac{(\psi / \chi)^{\lambda / 2}y^{\lambda-1}}{2 K_{\lambda} (\sqrt{\psi\chi})}\exp{\left\{ - \frac{\psi y+ \chi /y}{2}   \right\}},
\end{eqnarray}
where $\vecxi=(\varphi, \chi,\lambda)$.
Consider $Y$ and a random variable ${\bf V}\sim\mathcal{N}(\mathbf{0},{\matsig})$. Then, a generalized hyperbolic random variable ${\bf X}$, cf.\ \eqref{eqn:hyp}, can be generated via
\begin{eqnarray}
\label{hyp}
{ \bf X}={\bm \mu}+Y{\bm \alpha}+\sqrt{Y} {\bf V},
\end{eqnarray}
and it follows that ${\bf X}\mid Y\sim\mathcal{N}({\bm \mu}+y{\bm \alpha},y{\matsig})$.


Parameter estimation for the mixture of generalized hyperbolic distributions can be carried out via the expectation maximization (EM) algorithm \citep{DemLai77}. Note that the parameterization used in \eqref{eqn:hyp} requires the constraint $|{\matsig}|=1$ to ensure identifiability. Of course, this constraint is not practical for clustering and classification applications. Therefore, an alternative parameterization, setting $\omega =\sqrt{ \psi \chi}$ and $\eta =\sqrt {\chi / \psi}$, with $\eta=1$, is used. 
%
%
Under this parametrization, which is used by \cite{BroMcN15}, the density of the generalized hyperbolic distribution is 
\begin{equation}\label{eqn:hyp2} 
 f_{\text{H}}({\bf x}~|~{\bm \vartheta})=\left[\frac{\omega+ \delta  ({\bf x}, {\bm\mu}|{\matsig})}{\omega+{\bm \alpha}' {\bm\matsig}^{-1}{\bm \alpha}} \right]^{\frac{\lambda-\frac{p}{2}}{2}} \frac{K_{\lambda-\frac{p}{2}}\bigg(\sqrt {[\omega+{\bm\alpha}'{\bm\matsig}^{-1}{\bm \alpha}][\omega+\delta ({\bf x},{\bm\mu}|{\bm\matsig})]}\bigg)}
 {(2\pi)^{\frac{p}{2}}|{\bm\matsig}|^{\frac{1}{2}}K_\lambda(\omega)\exp{\{-({\bf x}-{\bm\mu})'{\bm\matsig}^{-1}{\bm\alpha}\}}}.\end{equation}
See \cite{BroMcN15} for further details. 

\section{Methodology}\label{sec:method}
\subsection{A Mixture of Generalized Hyperbolic Factor Analyzers}

Consider the number of free parameters in a $p$-dimensional, $G$-component mixture of generalized hyperbolic distributions. The scale matrices $\matsig_1,\ldots,\matsig_G$ contain $Gp(p+1)/2$ free parameters, i.e., a number that is quadratic in $p$; otherwise, the number of free parameters is linear in~$p$. For larger values of $p$, it is typically not viable to estimate $p(p+1)/2$ free parameters for each component scale matrix.
%
%
Introducing lower dimensional latent variables can help to resolve this problem. Given $p$-dimensional data ${\bf x}_1,\ldots,{\bf x}_n$, factor analysis finds uncorrelated $q$-dimensional latent factors ${\bf u}_1,\ldots,{\bf u}_n$ that explain a great deal of the variability in the data. 
The factor analysis model can be written 
\begin{eqnarray}
\label{hyp1}
{\bf X}_i={\bm \mu}+{\bm \Lambda} {\bf U}_{i}+ \vecepsilon_{i}, 
\end{eqnarray}
\noindent
for $i=1,\ldots,n$, where ${\bf U}_{i}\sim\mathcal{N}(\mathbf{0},{\bf I}_q)$, with $q\ll p$, and  $\vecepsilon_i\sim\mathcal{N}(\mathbf{0},{\bm \Psi_g})$. Note that ${\bf U}_{1},\ldots,{\bf U}_{n}$ are distributed independently, and independently of the errors $\vecepsilon_1,\ldots,\vecepsilon_n$, which are also distributed independently. The matrix $\bm\Lambda_g$ is a $p\times q$ matrix of factor loadings, and ${\bm \Psi_g}$ is a $p\times p$ diagonal matrix with strictly positive entries. The marginal distribution of $\vecX_i$ from model~\eqref{hyp1} is $\mathcal{N}(\bm \mu_g, \bm \Lambda_g \bm \Lambda_g' + \bm \Psi_g)$. \cite{ghahramani97} and \cite{mclachlan00a} consider a mixture of factor analyzers model, where
\begin{eqnarray}
\label{hyp1b}
{\bf X}_i={\bm \mu_g}+{\bm \Lambda}_g {\bf U}_{ig}+ \vecepsilon_{ig} \text{ with probability } \pi_g,
\end{eqnarray}
\noindent
for $i=1,\ldots,n$ and $g=1,\ldots, G$. 

To extend this model to the generalized hyperbolic distribution, first consider that  ${\bf V}$ in (\ref{hyp}) can be decomposed using a factor analysis model, i.e., $${\bf V}= {\bm\Lambda}{\bf U}+\vecepsilon,$$ 
where ${\bf U} \sim\mathcal{N}(\mathbf{0}, \mathbf{I}_q)$ and $\vecepsilon \sim\mathcal{N}({\bf 0}, {\bm \Psi})$. The resulting model can be represented as
\begin{eqnarray}
{\bf X}={\bm \mu}+Y{\bm \alpha}+\sqrt{Y}({\bm \Lambda  }{\bf U}+\vecepsilon),
\label{faMod}
\end{eqnarray}
%
and it follows that 
${\bf X}\mid y\sim\mathcal{N}({\bm \mu}+y{\bm \alpha},y({\bm \Lambda  \bm \Lambda'}+ {\bm \Psi}))$.
Then, in analogous fashion to the mixture of skew-$t$ factor analysis (MSTFA) model of \cite{murray14a}, we arrive at a mixture of generalized hyperbolic factor analyzers (MGHFA) model with density
\begin{equation}\label{eqn:model} 
 g({\bf x}~|~\pi_1,\ldots,\pi_g,{\bm \vartheta}_1,\ldots,{\bm \vartheta}_G)=\sum_{g=1}^G \pi_g f_{\text{H}}({\bf x}~|~{\bm \mu_g, {\bm \Lambda_g  \bm \Lambda_g'}+ {\bm \Psi_g},\bm \alpha_g, \lambda_g, \omega_g }).
\end{equation}

%
\subsection{Parameter Estimation}\label{sec:para}
Use $z_{ig}$ to denote component membership labels, where $z_{ig} =1$ if $\vecx_i$ is in component~$g$ and $z_{ig}=0$ otherwise, for $i=1, \ldots, n$ and $g=1, \ldots, G$. 
The alternating expectation-conditional maximization (AECM) algorithm \citep{MenVan97} can be useful when there are multiple sources of missing data and one wishes to find maximum likelihood estimates. The AECM algorithm is a variant of the EM algorithm and, like the EM algorithm, it is based on the complete-data log-likelihood, i.e., the likelihood of the observed data together with the unobserved (missing and / or latent) data. In our mixture of generalized hyperbolic factor analyzers model, the complete-data consist of the observed $\vecx_i$ as well as the missing labels $z_{ig}$, the latent $y_{ig}$, and the latent factors $\vecu_{ig}$. The AECM algorithm allows specification of different complete-data at each stage of the algorithm. 

In each E-step, the expected value of the complete-data log-likelihood is computed. As usual, the expected values of the $Z_{ig}$ are given by
\begin{equation*}
\e[Z_{ig}~|~\vecx_i]=\frac{\pi_gf_{\text{H}}({\bf x}_i~|~{\bm\vartheta_g})}{\sum_{h=1}^G\pi_hf_{\text{H}}({\bf x}_i~|~{\bm\vartheta_h})}=:\hat{z}_{ig}.
\end{equation*}
We also need the following expected values \cite[cf.][]{BroMcN15}: 
 \begin{equation}\begin{split}
&\e[Y_{ig}~|~\vecx_i,Z_{ig}=1]=\sqrt{\frac{\omega_g+ \delta  ({\bf x}_i, {\bm\mu_g}|{\matsig_g})}{\omega_g+{\bm \alpha_g}' {\bm\matsig_g}^{-1}{\bm \alpha_g}}}\nonumber\\
&\qquad\qquad\qquad\qquad\qquad\times\frac{K_{\lambda-\frac{p}{2}+1}\bigg(\sqrt {[\omega_g+{\bm\alpha_g}'{\bm\matsig_g}^{-1}{\bm \alpha_g}][\omega_g+\delta ({\bf x}_i,{\bm\mu_g}|{\bm\matsig_g})]}\bigg)}
{K_{\lambda-\frac{p}{2}}\bigg(\sqrt {[\omega_g+{\bm\alpha_g}'{\bm\matsig_g}^{-1}{\bm \alpha_g}][\omega_g+\delta ({\bf x}_i,{\bm\mu_g}|{\bm\matsig_g})]}\bigg)}\equalscolon a_{ig},\nonumber\\
&\e[1/Y_{ig}~|~\vecx_i,Z_{ig}=1]= -\frac{2\lambda_g-p}{\omega_g+ \delta  ({\bf x}_i, {\bm\mu_g}|{\matsig_g})}\\& \ \quad+\sqrt{\frac{\omega_g+{\bm \alpha_g}' {\bm\matsig_g}^{-1}{\bm \alpha_g}}{\omega_g+ \delta  ({\bf x}_i, {\bm\mu_g}|{\matsig_g})}}
\frac{K_{\lambda-\frac{p}{2}+1}\bigg(\sqrt {[\omega_g+{\bm\alpha_g}'{\bm\matsig_g}^{-1}{\bm \alpha_g}][\omega_g+\delta ({\bf x}_i,{\bm\mu_g}|{\bm\matsig_g})]}\bigg)}
{K_{\lambda-\frac{p}{2}}\bigg(\sqrt {[\omega_g+{\bm\alpha_g}'{\bm\matsig_g}^{-1}{\bm \alpha_g}][\omega_g+\delta ({\bf x}_i,{\bm\mu_g}|{\bm\matsig_g})]}\bigg)} \equalscolon b_{ig},\nonumber\\
\end{split}\end{equation}
\begin{equation}\begin{split}
&\e[\log Y_{ig}~|~\vecx_i,Z_{ig}=1]= \log\sqrt{\frac{\omega_g+ \delta  ({\bf x}_i, {\bm\mu_g}|{\matsig_g})}{\omega_g+{\bm \alpha_g}' {\bm\matsig_g}^{-1}{\bm \alpha_g}}}\nonumber\\
&\qquad\qquad\qquad+\frac{\partial}{\partial t} \log \left\{K_{t}\left(\sqrt {[\omega_g+{\bm\alpha_g}'{\bm\matsig_g}^{-1}{\bm \alpha_g}][\omega_g+\delta ({\bf x}_i,{\bm\mu_g}|{\bm\matsig_g})]} \right) \right\}\bigg|_{t=\lambda_g-\frac{p}{2}} \equalscolon c_{ig}.\nonumber
\end{split}\end{equation}
For convenience, set
 $n_g=\sum_{i=1}^n \hat z_{ig}$, $A_{g}=(1/n_g)\sum_{i=1}^n \hat z_{ig}a_{ig}$, $B_{g}=(1/n_g)\sum_{i=1}^n \hat z_{ig}b_{ig}$, and $C_{g}=(1/n_g)\sum_{i=1}^n \hat z_{ig}c_{ig}$.
When the latent factors $\vecU_{ig}$ are part of the complete-data, we will also need the following expectations: 
\begin{equation*}\begin{split}
&\e[{\bf U}_{ig}~|~\vecx_i, Z_{ig}=1]=\bm \beta_g({\bf x}_i-{\bm \mu_g}-a_{ig} \bm \alpha_g)\equalscolon \bm{E}_{1ig},\\
&\e[(1/Y_{ig}){\bf U}_{ig}~|~\vecx_i, Z_{ig}=1]=\bm\beta_g[b_{ig}({\bf x}_i-{\bm \mu_g})- \bm \alpha_g]\equalscolon \bm{E}_{2ig},\\
&\e[(1/Y_{ig}){\bf U}_{ig} {\bf U}_{ig}'~|~\vecx_i, Z_{ig}=1]=b_{ig}[{\bf I}_q - \bm \beta_g \bm \Lambda_g +\bm \beta_g ({\bf x}_i-{\bm \mu_g})({\bf x}_i-{\bm \mu_g})' \bm \beta_g']\\
&\qquad\qquad\qquad\qquad\qquad\quad-\bm \beta_g[ ({\bf x}_i-{\bm \mu_g})\vecalpha_g' +\vecalpha_g({\bf x}_i-{\bm \mu_g})' ]\bm\beta_g' + a_{ig}\bm\beta_g\vecalpha_g\vecalpha_g'\bm\beta_g'\equalscolon \bm{E}_{3ig},
\end{split}\end{equation*}
where $\bm \beta_g=\bm \Lambda_g'(\bm \Lambda_g\bm \Lambda_g'+\bm \Psi_g)^{-1}$.

At the first stage of the AECM algorithm, the complete-data comprise the observed $\vecx_i$, the missing labels $z_{ig}$, and the $y_{ig}$, and we update the mixing proportions $\pi_g$, the component means $\vecmu_g$, the skewness $\vecalpha_g$, the concentration $\omega_g$, and the index parameter $\lambda_g$. 
The complete-data log-likelihood is 
\begin{equation*}\begin{split}
\label{opt}
\log L_{1}=&\sum_{i=1}^n \sum_{g=1}^G z_{ig}[\log\pi_g+ \log\phi({\vecx_i}~|~{\bm \mu_g}+y_{ig} \bm \alpha_g, y_{ig}(\bm \Lambda_g \bm \Lambda_g'+\bm \Psi_g))+\log h(y_{ig}~|~\vecxi_g)].
\end{split}\end{equation*}
After forming the (conditional) expected value of $\log L_1$, we find that the updates for $\pi_g$, $\vecmu_g$, and $\vecalpha_g$ are given by:
\begin{eqnarray}
\hat\pi_g=\frac{n_g}{n},\qquad\hat\vecmu_g=\frac{\sum_{i=1}^n \hat{z}_{ig}\vecx_i(A_{g}b_{ig}-1)}{\sum_{i=1}^n  \hat{z}_{ig}(A_{g}b_{ig}-1)}, 
\qquad\text{and}\qquad\hat\vecalpha_g=\frac{\sum_{i=1}^n  \hat{z}_{ig}\vecx_i(b_{ig}-B_g)}{\sum_{i=1}^n  \hat{z}_{ig}(A_{g}b_{ig}-1)}, \nonumber
\end{eqnarray}
respectively.
The parameters $\omega_g$ and $\lambda_g$ are estimated by maximizing the following function,
$$q_g(\omega_g, \lambda_g)=-\log K_\lambda(\omega_g)+(\lambda_g -1)C_g- \frac{\omega_g}{2}(A_g+B_g),$$
and the associated updates are:
\begin{equation*}\begin{split}
&\hat{\lambda}_g = C_g \hat{\lambda}_g^{\prev}\left[ \frac{ \partial }{ \partial t }  \log K_t \left(\hat{\omega}_g^{\prev}\right) \Big|_{t=\hat{\lambda}_g^{\prev}} \right]^{-1},\\
&\hat{\omega}_g = \hat{\omega}_g^{\prev} - \left[{ \left. \frac{\partial }{\partial t  }  q_g\big( t,\hat{\lambda}_g\big) \right|_{t = \hat{\omega}_g^{\prev}}}\right]\left[{ \left. \frac{\partial^2 }{\partial t^2 } q_g\big(t,\hat{\lambda}_g\big) \right|_{t = \hat{\omega}_g^{\prev}}}\right]^{-1},
\end{split}\end{equation*}
where the superscript `prev' is used to distinguish the previous estimate from the current one --- see \cite{BroMcN15} for details.

At the second stage of the AECM algorithm, the complete-data comprise the observed $\vecx_i$, the missing labels $z_{ig}$, the $y_{ig}$, and the $\vecu_{ig}$. At this stage, we update $\bm\Lambda_g$ and $\bm\Psi_g$, and the complete-data log-likelihood can be written
\begin{equation*}\begin{split}
\label{opt}
\log L_2 &=\sum_{i=1}^n \sum_{g=1}^G z_{ig}\big[\log\pi_g+
\log\phi({\vecx_i}~|~{\bm \mu_g}+y_{ig} \bm \alpha_g + \bm \Lambda_g \vecu_{ig}, y_{ig} \bm \Psi_g)\\
&\qquad\qquad\qquad\qquad\qquad\qquad\qquad\qquad+
\log\phi({\vecu_{ig}}~|~{\bm 0},y_{ig} \bm I_q)+
\log h(y_{ig}~|~\vecxi_g)\big]\\
&=C-\frac{1}{2}\sum_{i=1}^n \sum_{g=1}^Gz_{ig}\log|\noisev_g|-\frac{1}{2}\sum_{i=1}^n \sum_{g=1}^Gz_{ig}\bigg[\frac{1}{y_{ig}}\tr\{(\vecx_i-\vecmu_g)(\vecx_i-\vecmu_g)'\noisev_g^{-1}\}\\&\qquad-2\tr\{(\vecx_i-\vecmu_g)\vecalpha_g'\noisev_g^{-1}\}+y_{ig}\tr\{\vecalpha_g\vecalpha_g'\noisev_g^{-1}\}-\frac{2}{y_{ig}}\tr\{(\vecx_i-\vecmu_g)'\noisev_g^{-1}\load_g\vecu_{ig}\}\\
&\qquad +2\tr\{\vecalpha_g'\noisev_g^{-1}\load_g\vecu_{ig}\}+\frac{1}{y_{ig}}\tr\{\load_g\vecu_{ig}\vecu_{ig}'\load_g'\noisev_g^{-1}\}\bigg],\\
\end{split}\end{equation*}
where $C$ is constant with respect to $\bm \Lambda_g$ and $\bm \Psi_g$.
As shown in in Appendix~\ref{appendix}, the updates, which follow from the (conditional) expected value of $\log L_2$, are given by:
\begin{equation*}\begin{split}
\hat\load_g &= \bigg\{\sum_{i=1}^n\hat{z}_{ig}\left[(\vecx_i-\hat\vecmu_g)\bm{E}_{2ig}'-\hat\vecalpha_g\bm{E}_{1ig}'\right]\bigg\}\bigg\{\sum_{i=1}^n\hat{z}_{ig}\bm{E}_{3ig}\bigg\}^{-1},\\
\hat\noisev_g &= \frac{1}{n_g}\text{diag}\bigg\{\sum_{i=1}^n\hat{z}_{ig}\big[b_{ig}(\vecx_i-\hat\vecmu_g)(\vecx_i-\hat\vecmu_g)'-2\hat\vecalpha_g(\vecx_i-\hat\vecmu_g)'+a_{ig}\hat\vecalpha_g\hat\vecalpha_g'\\
&\qquad\qquad\qquad\qquad\qquad-2(\vecx_i-\hat\vecmu_g)\bm{E}_{2ig}'\hat\load_g'+2\hat\vecalpha_g\bm{E}_{1ig}'\hat\load_g'+\hat\load_g\bm{E}_{3ig}\hat\load_g'\big]\bigg\}.
\end{split}\end{equation*}

Note that the inversion of the matrix $(\bm\Lambda_g\bm\Lambda_g'+\bm\Psi_g)$ requires the inversion of a $p \times p$ matrix, which can be slow for larger values of $p$. The Woodbury identity \citep{woodbury50} gives the formula
\begin{eqnarray}
(\bm\Lambda_g \bm\Lambda_g'+\bm\Psi_g)^{-1} =\bm\Psi_g^{-1}-\bm\Psi_g^{-1}\bm\Lambda_g (\ident_q+\bm\Lambda_g'\bm\Psi_g^{-1}\bm\Lambda_g)^{-1}\bm\Lambda_g'\bm\Psi_g^{-1},
\label{inversion}
\end{eqnarray}
\noindent
which requires the inversion of diagonal $p\times p$ matrices and a $q \times q$ matrix, resulting in a significant speed-up when $q\ll p$.

The algorithm can be initialized using randomly generated weights $\hat{z}_{ig}$. One of the most commonly used initialization techniques in the literature is $k$-means clustering, and many clustering packages for {\sf R} \citep{R14} use this method as a default starting criterion \citep[e.g.,][]{berge13,lee13b,mcnicholas11}. 

\subsection{Model Selection and  Convergence}\label{subs:modsel}
In addition to parameter estimation, we need to choose the number of components $G$ and the number of factors $q$. The Bayesian information criterion \cite[BIC;][]{Sch78} is used to select $G$ and $q$, and is defined as 
$\text{BIC}=2l(\vecx~|~\hat\varthet)-\rho\log n$, where $l(\vecx~|~ \varthet)$ is the maximized log-likelihood, $\hat\varthet$ is the vector of parameters that maximize the log-likelihood, $\rho$ is the number of free parameters, which is $(G-1)+G[3p+2+pq-q(q-1)/2]$ for the MGHFA model, and $n$ is the number of units. The BIC is often used for model selection in model-based clustering, and arguments for its use in this context are given by \cite{CamFra97} and \cite{DasRaf98}.
Simulation results supporting the use of the BIC for selecting the number of factors in a factor analysis model are given by \cite{LopWes04}. More recent accounts of the advantages and drawbacks of the BIC and some alternatives are given by \cite{maugis09}, \cite{henning2010}, \cite{wei14}, and \cite{bhattacharya14}. A simulation study on the sensitivity of the BIC to the number of components is presented in Section~\ref{sec:simulation}.

For the analyses reported herein, convergence of the AECM algorithm is determined using an approach based on the Aitken acceleration \citep{ait26}. Let $l^{(k)}$ be the value of the log-likelihood at the iteration $k$. The Aitken acceleration is given by 
\begin{eqnarray*}
a^{(k)}=\frac{l^{(k+1)}-l^{(k)}}{l^{(k)}-l^{(k-1)}},
\end{eqnarray*}
and an asymptotic estimate of the log-likelihood at iteration $k+1$ is
 \begin{eqnarray*}
l^{(k+1)}_\infty=l^{(k)}+\frac{1}{1-a^{(k)}}(l^{(k+1)}-l^{(k)}).
\end{eqnarray*}
The algorithm can be considered to have converged if $l^{(k)}_{\infty}-l^{(k)}< \epsilon$, where $\epsilon$ is a positive small value \citep{BohDie94,Lin95}.

\subsection{Model-Based Classification}

Model-based classification is a semi-supervised analogue of model-based clustering. If we suppose that $k$ of the $n$ samples are labelled, then we can use model-based classification to classify the other $n-k$ samples within a joint likelihood framework. Ordering the $n$ samples so that it is the first $k$ that are labelled, the model-based classification likelihood for the generalized hyperbolic factor analyzers model is
\begin{eqnarray}
\label{classifopt}
L_{\text{class}}=\prod_{i=1}^k\prod_{g=1}^G\pi_gf_{\text{H}}(\vecx_i~|~\varthet_g)\prod_{j=k+1}^n\sum_{h=1}^H\pi_hf_{\text{H}}(\vecx_j~|~\varthet_h),
\end{eqnarray}
where $H\geq G$. Note that it is very often assumed that $H=G$.
Parameter estimation for model-based classification proceeds in a similar fashion to model-based clustering; see \cite{mcnicholas10b} for details in the case of the mixture of factor analyzers model and parsimonious extensions thereof. 

\section{Applications}\label{sec:data}
In this section, we illustrate the MGHFA for model-based clustering and classification using real data sets. The software we use for the analyses herein is available via the \texttt{MixGHD} package \citep{tortora14} for {\sf R}. We compare our method with its Gaussian analogue, i.e., the mixture of factor analyzers model (MFA), which is fitted using the \texttt{pgmm} package \citep{mcnicholas11} for {\sf R}, and the MSTFA model. We also compare with the mixture of high-dimensional Gaussian mixture models (HD-GMM), which performed well in a survey of mixture model-based approaches for high-dimensional data \citep[cf.][]{Bouveyron13}. This method is available via the \texttt{HDclassif} package \citep{berge13} for {\sf R}. MSTFA and HD-GMM are families of models and, to facilitate comparison, we will use the most general model in each case. As a fourth competitor, we use an approach based on the famous $k$-means clustering method. 
Several extensions of $k$-means clustering for high-dimensional data sets have recently been proposed \citep{Vickie01,rocGat11,TimCeu13}; among them, factorial k-means (FKM) is available in the \texttt{clustrd} package \citep{markos13} for {\sf R}. We use FKM in our clustering comparisons, but it cannot be used for classification.

We use the adjusted Rand index \cite[ARI;][]{HubAra85} to evaluate the classification performance of the methods. The ARI is the Rand index \citep{rand71} corrected for agreement by chance. The ARI is equal to $1$ when there is perfect class agreement, and its expected value under random classification is $0$.

In the following real and simulated examples, all of the methods have been initialized using the $k$-means algorithm. For each approach, we used twenty different starts and reported the result from the best model.

\subsection{Model-Based Clustering}
\subsubsection{Italian Wine Data}
These data are the result of physical and chemical analysis of wines grown in the same region in Italy but derived from three different cultivars: Barolo, Grignolino, and Barbera \citep{ForArm86}. There are 178 samples of 27 physical and chemical measurements available in the \texttt{pgmm} package for {\sf R}. 
We set the number of components $G=3$ and the number of factors $q$ is selected based on the BIC. A $q=2$ factor model is selected, and the associated MGHFA model gives very good classification performance (Table~\ref{Winemiss1}; $\text{ARI}=0.800$).
Our approach is compared with the MSTFA, MFA, HD-GMM, and FKM approaches. The BIC criterion selects $q=2$ latent factors for MSTFA and HD-GMM; however, the associated models give relatively poor classification performance, with ARI values of $0.741$ and $0.284$, respectively (Table~\ref{Winemiss}).   The BIC selects  $q=3$ for MFA and it gives a better classification performance than MSTFA or HD-GMM ($\text{ARI}=0.792$).
Note that the number of factors for FKM is fixed at $q=G-1$, as suggested by the authors, and the classification results are similar to what would be expected from a random classification ($\text{ARI} \approx 0$).
\begin{table*}[t]
\begin{center}
\caption{\label{Winemiss1} Cross-tabulation of the true versus predicted class labels for model-based clustering on the wine data set using MGHFA.}
\begin{tabular*}{1.00\textwidth}{@{\extracolsep{\fill}}lccc}
\hline
&\multicolumn{3}{c}{MGHFA}\\
\cline{2-4}
&1 &2&3 \\
\hline
Barolo&59 &0&0  \\
Grignolino&10 &60&1\\
Barbera&0 &1&47 \\
\hline
\end{tabular*}
\end{center}
\end{table*}

\begin{table*}[t]
\begin{center}
\caption{\label{Winemiss} Cross-tabulation of the true versus predicted class labels for model-based clustering on the wine data using MSTFA, MFA, HD-GMM, and FKM, respectively.}
\begin{tabular*}{1.00\textwidth}{@{\extracolsep{\fill}}lcccrccc}
\hline
&\multicolumn{3}{c}{MSTFA}&&\multicolumn{3}{c}{MFA}\\
\cline{2-4}\cline{6-8}
&1 &2&3 &&1 &2&3\\
\hline
Barolo&58 &0&1 &&59 &0&0\\
Grignolino&0 &65&6&&4 &58&9\\
Barbera&0 &11&37 &&0&0&48\\
\hline
\hline
&\multicolumn{3}{c}{HD-GMM}&&\multicolumn{3}{c}{FKM}\\
\cline{2-4}\cline{6-8}
&1 &2&3 &&1 &2&3\\
\hline
Barolo&17 &9&33 &&23 &14&22 \\
Grignolino&18 &20&33&&18&33&20\\
Barbera&4 &7&37&&18&13&17\\
\hline
\end{tabular*}
\end{center}
\end{table*}

\subsubsection{Liver Data}
\cite{chen02} study gene expression microarray data to compare patients with hepatocellular carcinoma (HCC) to those with healthy (non-tumour) liver tissue. The data contain 179 samples --- 104 from patients with HCC and 75 non-tumour samples --- for 85 genes.
We set $G=2$ components for the MGHFA, MSTFA,  and HD-GMM models, and the BIC selects $q=2$ latent factors. The associated MGHFA model gives good classification performance, misclassifying only 16 of the 179 samples (Table~\ref{Chengmiss}, $\text{ARI}=0.672$). The MSTFA, MFA, HD-GMM, and FKM approaches give worse results, with 80, 80, 80, and 85 misclassified samples, respectively (ARI $\approx 0$).
\begin{table*}
\begin{center}
\caption{\label{Chengmiss} Cross-tabulation of the true versus predicted class labels for model-based clustering on the liver data set using MGHFA, MFA, HD-GMM, and FKM, respectively.}
\begin{tabular*}{1.00\textwidth}{@{\extracolsep{\fill}}lcccccccccc}
\hline
&\multicolumn{2}{c}{MGHFA}&\multicolumn{2}{c}{MSTFA}&\multicolumn{2}{c}{MFA}&\multicolumn{2}{c}{HD-GMM}&\multicolumn{2}{c}{FKM}\\
\cline{2-3}\cline{4-5}\cline{6-7}\cline{8-9}\cline{10-11}
&1 &2&1 &2&1 &2&1 &2&1 &2\\
\hline
HCC &97 &7  &13&91 &13 &91&13 &91  &52 &52  \\
Non-tumour &9 &66 &8&67&8 &67&8 &67 &33 &42\\
\hline

\end{tabular*}
\end{center}
\end{table*}

\subsection{Model-Based Classification}
\subsubsection{Sonar Data}\label{sec:sonar}
\cite{GorSej88} report the patterns obtained by bouncing sonar signals at various angles and under various conditions. There are 208 patterns in all, 111 obtained by bouncing sonar signals off a metal cylinder and 97 obtained by bouncing signals off rocks. Each pattern is a set of 60 numbers (variables) taking values between~0 and~1. These data are sourced from the UCI machine learning repository. We use a familiar approach \citep[cf.][]{mcnicholas10b} to designate some of the patterns as unlabelled. Specifically, 
a probability is specified \textit{a priori}: $0.3$ in this case. Then, for each $\vecx_i$, a random number~$r_i$ is generated from a uniform distribution on [0,1]. If $r_i<0.3$, then~$\vecx_i$ is taken as unlabelled; otherwise,~$\vecx_i$ is taken as labelled. Applying this approach to the sonar data leads to a data set where 54 of the 208 patterns are unlabelled.

%
Applying the MGHFA model to this data set leads to the selection of a model with $q=2$ factors. The results are compared with HD-GMM,  MSTFA, and MFA. 
The MGHFA, MSTFA, and HD-GMM approaches give reasonable classification performance, each with an associated ARI value of $0.339$ (Table~\ref{Sonarmiss}). This classification performance is better than the MFA model ($\text{ARI}=0.217$). Although no approach gives particularly good classification performance, these data are notoriously difficult to classify. To put this point in context, we can compare our results to those of \cite{TanDow05}, who report  classification accuracy of $76.0 \pm 9.2$ in an analysis where 10\% of the patterns are unlabelled. The MGHFA, MSTFA, and HD-GMM approaches give 79.6\% classification accuracy (Table~\ref{Sonarmiss}) in an analysis where 26\% of the patterns are unlabelled.
\begin{table*}
\begin{center}
\caption{\label{Sonarmiss} Cross-tabulation of the true versus predicted class labels, for the unlabelled observations, for model-based classification of the sonar data set using MGHFA, MSTFA, MFA, and HD-GMM respectively.}
\begin{tabular*}{1.00\textwidth}{@{\extracolsep{\fill}}lcccccccc}
\hline
&\multicolumn{2}{c}{MGHFA} & \multicolumn{2}{c}{MSTFA}& \multicolumn{2}{c}{MFA}& \multicolumn{2}{c}{HD-GMM}\\
\cline{2-3}\cline{4-5}\cline{6-7}\cline{8-9}
&1 &2&1 &2&1 &2&1 &2\\
\hline
Rock &18 &5 &18& 5&21 &2& 15& 8\\
Metal&6&25 &6&25&12&19&3 &28\\
\hline
\end{tabular*}
\end{center}
\end{table*}

\subsubsection{Italian Olive Oil Data}

\cite{forina82a} and \cite{forina82b} report the eight fatty acids found by lipid fraction of 572 italian olive oils. The oils come from three regions of Italy --- Southern Italy, Sardinia, and Northern Italy --- and these regions can be further divided into nine different areas. The data are available in the  \texttt{pgmm} package for {\sf R}. We use the same procedure as in Section~\ref{sec:sonar} to generate a data set where 171 of the 572 oils are taken to be unlabelled. The selected number of factors for the MGHFA model is $q=2$ and, as for the MSTFA and MFA models, perfect classification results are obtained for the three regions as well as very good results ($\text{ARI}\approx 0.91$) for the nine areas (Table~\ref{ARI}). The HD-GMM approach gives similar, but slightly inferior, classification performance on these data.
\begin{table*}
\begin{center}
\caption{\label{ARI} ARI values, based on predicted classifications for the unlabelled observations, for the MGHFA, MSTFA, MFA, and HD-GMM models for model-based classification on the olive oil data.}
\begin{tabular*}{1.00\textwidth}{@{\extracolsep{\fill}}lrrrr}
\hline
&{MGHFA}&MSTFA& MFA&HD-GMM\\
\hline
By regions ($G=3$) &$1$ & $1$& $1$&$0.800$\\
By areas ($G=9$) &$0.913$ &$0.911$ & $0.906$& $0.875$\\
\hline
\end{tabular*}
\end{center}
\end{table*}

\section{Simulation studies}\label{sec:simulation}

In addition to the real data applications of Section~\ref{sec:data}, our MGHFA model and the same comparator approaches are applied to simulated data. We consider data from three different mixture distributions: a mixture of Gaussian distributions, a mixture of skew-normal distributions, and a mixture of generalized hyperbolic distributions. 
Within each mixture component, $n_g $ $p$-dimensional vectors $\vecx_i$ are generated.  The mixing proportions $\pi_g$ are equal across components. 
Each mixture component is centred on a different point, with the locations uniformly distributed on a hypercube of side length 200. The $p\times p$ covariance / scale matrices are generated by first generating an identity matrix and then adding diagonal elements in the interval $[-0.9, 0.9]$. When applicable, the skewness parameter $\bm \alpha_g$ is randomly generated in $\pm[10,20]$, and the values of the other parameters for the mixture of generalized hyperbolic distributions are $ \omega=1, \lambda=0.5$. In the following simulation study, we consider such mixtures with $G=2,3$ and $p=10,100,500$. The Gaussian data sets are generated via the {\sf R} function \texttt{rnorm}, the skew-normal data using the \texttt{rdmsn} function from the \texttt{EMMIXskew} package \citep{Wang13} for {\sf R}, and the generalized hyperbolic data using our own code.

For clustering, all five methods give good performance when dealing with data from Gaussian and skew-normal mixtures; however,  FKM performs poorly when $G=3$ (Table~\ref{Sim}). As one would expect, MGHFA works better than its competitors when the data are generated from generalized hyperbolic mixtures.
\begin{table}[!h]
{\footnotesize
\begin{center}
\caption{\label{Sim} ARI values for the MGHFA, MSTFA, MFA, HD-GMM, and FKM models for clustering on simulated data.}
\begin{tabular*}{1.00\textwidth}{@{\extracolsep{\fill}}lcccccccc}
\hline
Data &$p$ & $G$& $n$&{MGHFA} &MSTFA & {MFA}&{HD-GMM}&FKM\\
\hline
Gaussian& 10&2 &200 &1 & 1&  1& 1&1\\
Gaussian& 100&2 &200&1 &1&1 & 1&1\\
Gaussian& 500&2  &200&1&1&  1& 1&1\\
\hline
Gaussian& 10&3  &300&1 &1 & 1& 1&0\\
Gaussian& 100&3&300 &1 & 1&1 & 1&0.12\\
Gaussian& 500&3 &300 &1& 1& 1& 1&0.12\\
\hline
\hline
Skew-normal& 10&2 &200 &1 &1&1&1&0\\
Skew-normal& 100&2 &200&1 &1 &1 &1&1 \\
Skew-normal& 500&2&200&  1&1& 1&1& 1\\
\hline
Skew-normal& 10&3  &300&1 &1&   1&1&0\\
Skew-normal& 100&3 &300&1 & 1&1 &1&0.21\\
Skew-normal& 500&3  &300&1&1 & 1&1& 0.21\\
\hline
\hline
GHD& 10&2 &200 &1 &1&1&1 &0.02\\
GHD& 100&2&200 & 1&1& 1 & 1&0.06\\
GHD& 500&2&200&  1&0.72 & 0.96& 0.06&0.70\\
\hline
GHD& 10&3 &300 &0.99& 1 &0.92& 0.92&0.02\\
GHD& 100&3 &300&1 &0.91& 1 & 1&0.06\\
GHD& 500&3&  300&1&0.72& 1& 1&0.81\\
\hline
\hline
\end{tabular*}
\end{center}}
\end{table}

For classification, we randomly select $30\%$ of the points and treat them as unlabelled. All four methods gave perfect or near perfect performance for all cases  (Table \ref{Sim2}).
\begin{table}[!h]
{\footnotesize
\begin{center}
\caption{\label{Sim2} ARI values for the MGHFA, MSTFA, MFA, HD-GMM, and FKM models for classification on the simulated data sets.}
\begin{tabular*}{1.00\textwidth}{@{\extracolsep{\fill}}lccccccc}
\hline
Data &$p$ & $G$& $n$&{MGHFA} & MSTFA& {MFA}&{HD-GMM}\\
\hline
Gaussian& 10&2&200  &1 &1&   1& 1\\
Gaussian& 100&2&200 &1 &1& 1 & 1\\
Gaussian& 500&2&200&  1&1&  1& 1\\
\hline
Gaussian& 10&3 &300 &1 &0.82&   1& 1\\
Gaussian& 100&3&300 &1 &1& 1 & 1\\
Gaussian& 500&3&300&  1&1&  1& 1\\
\hline
\hline
Skew-normal& 10&2&200  &1 &1&1&1\\
Skew-normal& 100&2&200 &1 &1& 1 &1 \\
Skew-normal& 500&2&200&  1&1&1&1\\
\hline
Skew-normal& 10&3  &300&1 &1&   1&1\\
Skew-normal& 100&3 &300&1 &0.91& 1 &1\\
Skew-normal& 500&3&300&  1&1& 1&1\\
\hline
\hline
GHD& 10&2  &200&1 &1&0.80&0.80 \\
GHD& 100&2 &200& 1& 1&1 & 1\\
GHD& 500&2&200&  1&1&  1& 1\\
\hline
GHD& 10&3 &300 &1&1& 0.93& 0.93\\
GHD& 100&3 &300&1 &1& 1 & 1\\
GHD& 500&3&300&  1&1 &1& 1\\
\hline
\hline
\end{tabular*}
\end{center}}
\end{table}

%
Finally, we conduct a simulation to study the performance of the BIC in model selection. In Section \ref{subs:modsel}, we suggest using the BIC to select the number of components; however, many authors \cite[e.g.,][]{henning2010,bhattacharya14} have pointed out that the BIC may suggest a larger number of mixture components than what seems to be a reasonable number of clusters. For this reason, we carry out an empirical study for the sensitivity of the BIC to the number of components for each model-based method used in our analyses. Specifically, using the same scheme as in Section \ref{sec:simulation}, we consider data from three different mixture distributions: a mixture of Gaussian distributions, a mixture of skew-normal distributions, and a mixture of generalized hyperbolic distributions, with $p=10$ and $G=\{ 2,3,4,5\}$. We apply each method, i.e., MGHFA, MSTFA, MFA, and HD-GMM, with $G\in[2,10]$. Table~\ref{BicSim} shows the selected number of components according to the BIC. For each approach, we used twenty different starts and reported the result from the best model. When using the MGHFA, the BIC always finds the true number of components. When using methods based on the Gaussian distribution, the BIC fails to detect the true number of components on non-Gaussian distributed clusters; specifically, it overestimates the number of components, as expected. In fact, the methods based on the Gaussian distribution also overestimate the number of components in some cases where the data are generated from Gaussian mixtures. 
\begin{table}[!h]
{\footnotesize
\begin{center}
\caption{\label{BicSim}The number of selected components using the BIC for the MGHFA,  MSTFA, MFA, and HD-GMM models on the final simulation study.}
\begin{tabular*}{1.00\textwidth}{@{\extracolsep{\fill}}lccccccc}
\hline
Data &$p$ & $G$& $n$&{MGHFA} &MSTFA & {MFA}&{HD-GMM}\\
\hline
Gaussian& 10&2 &200 &2 &2&  2& 2\\
Gaussian& 10&3 &300&3&3&3& 3\\
Gaussian& 10&4  &400&4&6&  5& 4\\
Gaussian& 10&5&500 &5&5&  6&4 \\
\hline
\hline
Skew-normal& 10&2 &200 &2 &2&2&3\\
Skew-normal& 10&3 &300& 3&3 &3 & 4\\
Skew-normal& 10&4&400& 4 &5& 5&4\\
Skew-normal& 10&5&500 &5&5&6&7\\
\hline
GHD& 10&2 &200 &2 &2&4&8 \\
GHD& 10&3&300 & 3&3& 6& 9\\
GHD& 10&4&400&  4&4&8& 10\\
GHD& 10&5&500&  5& 5& 8&10\\
\hline
\hline
\end{tabular*}
\end{center}}
\end{table}

\section{Summary}\label{sec:conc}
The MFA model has been extended to the generalized hyperbolic distribution. Parameter estimation was outlined via an AECM algorithm and the BIC was used for model selection. Our MGHFA model was applied to real and simulated data for clustering and classification, where its performed favourably when compared to several other approaches. Looking forward, further parsimony can be achieved by considering a generalized hyperbolic analogue of the family of models introduced by \cite{McNicholas08,McNicholas10}. It will also be interesting to consider an analogue of LASSO-penalized BIC that \cite{bhattacharya14} used for the mixture of factor analyzers model and a family of Gaussian mixture models based thereon. Variational Bayes approximations will be investigated as an alternative to the AECM algorithm for parameter estimation \citep[cf.][]{subedi14}. Finally, trimming approaches will be considered \citep[cf.][]{ritter14}.

\appendix
\section{Updates for component covariance parameters}\label{appendix}

At the second stage for our AECM algorithm, the (conditional) expected value of complete-data log-likelihood is given by
\begin{equation*}\begin{split}
\label{opt}
{Q}_2 &=C-\frac{1}{2}\sum_{i=1}^n \sum_{g=1}^G\hat{z}_{ig}\log|\noisev_g|-\frac{1}{2}\sum_{i=1}^n \sum_{g=1}^G\hat{z}_{ig}\bigg[b_{ig}\tr\{(\vecx_i-\hat\vecmu_g)(\vecx_i-\hat\vecmu_g)'\noisev_g^{-1}\}
-2\tr\{(\vecx_i-\hat\vecmu_g)\hat\vecalpha_g'\noisev_g^{-1}\}\\&
\quad+a_{ig}\tr\{\hat\vecalpha_g\hat\vecalpha_g'\noisev_g^{-1}\}-2\tr\{(\vecx_i-\hat\vecmu_g)'\noisev_g^{-1}\load_g\bm{E}_{2ig}\} +2\tr\{\hat\vecalpha_g'\noisev_g^{-1}\load_g\bm{E}_{1ig}\}+\tr\{\load_g\bm{E}_{3ig}\load_g'\noisev_g^{-1}\}\bigg],\\
\end{split}\end{equation*}
where $C$ is constant with respect to $\bm \Lambda_g$ and $\bm \Psi_g$.
Differentiating ${Q}_2$ with respect to $\bm \Lambda_g$ gives
\begin{equation*}\begin{split}
S_1(\load_g,\noisev_g)=\frac{\partial{Q}_2}{\partial \bm \Lambda_g}=
-\frac{1}{2}\sum_{i=1}^n\hat{z}_{ig}\left[-2\noisev_g^{-1}(\vecx_i-\hat\vecmu_g)\bm{E}_{2ig}'+2\noisev_g^{-1}\hat\vecalpha_g\bm{E}_{1ig}'
+\noisev_g^{-1}\load_g(\bm{E}_{3ig}'+\bm{E}_{3ig})\right]\end{split}\end{equation*}
Note that $\bm{E}_{3ig}$ is a symmetric matrix. Now, solving $S_1(\hat\load_g,\noisev_g)=\bm{0}$ gives the update:
$$\hat\load_g = \bigg\{\sum_{i=1}^n\hat{z}_{ig}\left[(\vecx_i-\hat\vecmu_g)\bm{E}_{2ig}'-\hat\vecalpha_g\bm{E}_{1ig}'\right]\bigg\}\bigg\{\sum_{i=1}^n\hat{z}_{ig}\bm{E}_{3ig}\bigg\}^{-1}.$$
Differentiating ${Q}_2$ with respect to $\noisev_g^{-1}$ gives
\begin{equation*}\begin{split}
S_2(\load_g,\noisev_g)=\frac{\partial{Q}_2}{\partial \noisev_g^{-1}}=
\frac{1}{2}\sum_{i=1}^n\hat{z}_{ig}\noisev_g
-\frac{1}{2}\sum_{i=1}^n\hat{z}_{ig}\big[&b_{ig}(\vecx_i-\hat\vecmu_g)(\vecx_i-\hat\vecmu_g)'-2\hat\vecalpha_g(\vecx_i-\hat\vecmu_g)'+a_{ig}\hat\vecalpha_g\hat\vecalpha_g'\\
&-2(\vecx_i-\hat\vecmu_g)\bm{E}_{2ig}'\load_g'+2\hat\vecalpha_g\bm{E}_{1ig}'\load_g'+\load_g\bm{E}_{3ig}\load_g'\big].\end{split}\end{equation*}

Now, solving $\text{diag}\{S_2(\hat\load_g,\hat\noisev_g)\}=\bm{0}$ gives the update:
\begin{equation*}\begin{split}
\hat\noisev_g &= \frac{1}{n_g}\text{diag}\bigg\{\sum_{i=1}^n\hat{z}_{ig}\big[b_{ig}(\vecx_i-\hat\vecmu_g)(\vecx_i-\hat\vecmu_g)'-2\hat\vecalpha_g(\vecx_i-\hat\vecmu_g)'+a_{ig}\hat\vecalpha_g\hat\vecalpha_g'\\
&\qquad\qquad\qquad\qquad\qquad-2(\vecx_i-\hat\vecmu_g)\bm{E}_{2ig}'\hat\load_g'+2\hat\vecalpha_g\bm{E}_{1ig}'\hat\load_g'+\hat\load_g\bm{E}_{3ig}\hat\load_g'\big]\bigg\}.\end{split}\end{equation*}

\bibliographystyle{chicago}

\begin{thebibliography}{}

\bibitem[\protect\citeauthoryear{Aitken}{Aitken}{1926}]{ait26}
Aitken, A. (1926).
\newblock On {B}ernoulli's numerical solution of algebraic equations.
\newblock {\em Proceedings of the Royal Society of Edimburgh\/}~{\em 46},
  289--305.

\bibitem[\protect\citeauthoryear{Andrews and McNicholas}{Andrews and
  McNicholas}{2012}]{andrews12}
Andrews, J.~L. and P.~McNicholas (2012).
\newblock Model-based clustering, classification, and discriminant analysis via
  mixtures of multivariate $t$-distributions.
\newblock {\em Statistics and Computing\/}~{\em 22\/}(5), 1021--1029.

\bibitem[\protect\citeauthoryear{Andrews and McNicholas}{Andrews and
  McNicholas}{2011a}]{andrews11b}
Andrews, J.~L. and P.~D. McNicholas (2011a).
\newblock Extending mixtures of multivariate t-factor analyzers.
\newblock {\em Statistics and Computing\/}~{\em 21\/}(3), 361--373.

\bibitem[\protect\citeauthoryear{Andrews and McNicholas}{Andrews and
  McNicholas}{2011b}]{andrews11c}
Andrews, J.~L. and P.~D. McNicholas (2011b).
\newblock Mixtures of modified t-factor analyzers for model-based clustering,
  classification, and discriminant analysis.
\newblock {\em Journal of Statistical Planning and Inference\/}~{\em 141\/}(4),
  1479--1486.

\bibitem[\protect\citeauthoryear{Baek, McLachlan, and Flack}{Baek
  et~al.}{2010}]{Baek10}
Baek, J., G.~J.~M. McLachlan, and L.~Flack (2010).
\newblock Mixtures of factor analyzers with common factor loadings:
  Applications to the clustering and visualization of high-dimensional data.
\newblock {\em IEEE Transactions on Pattern Analysis and Machine
  Intelligence\/}~{\em 32\/}(7), 1298--1309.

%
\bibitem[\protect\citeauthoryear{Barndorff-Nielsen and
  Halgreen}{Barndorff-Nielsen and Halgreen}{1977}]{barndorff77}
Barndorff-Nielsen, O. and C.~Halgreen (1977).
\newblock Infinite divisibility of the hyperbolic and generalized inverse
  {G}aussian distributions.
\newblock {\em Z.\ Wahrscheinlichkeitstheorie Verw.\ Gebiete\/}~{\em 38},
  309--311.


\bibitem[\protect\citeauthoryear{Berg\'e, Bouveyron, and Girard}{Berg\'e
  et~al.}{2013}]{berge13}
Berg\'e, L., C.~Bouveyron, and S.~Girard (2013).
\newblock Hdclassif: High dimensional supervised classification and clustering.
\newblock R package version 1.2.2.

\bibitem[\protect\citeauthoryear{Bhattacharya and McNicholas}{Bhattacharya and
  McNicholas}{2014}]{bhattacharya14}
Bhattacharya, S. and P.~D. McNicholas (2014).
\newblock A {LASSO}-penalized {BIC} for mixture model selection.
\newblock {\em Advances in Data Analysis and Classification\/}~{\em 8\/}(1),
  45--61.

\bibitem[\protect\citeauthoryear{Bl{\ae}sild}{Bl{\ae}sild}{1978}]{blaesild78}
Bl{\ae}sild, P. (1978).
\newblock The shape of the generalized inverse {G}aussian and hyperbolic
  distributions.
\newblock Research Report~37, Department of Theoretical Statistics, Aarhus
  University, Denmark.

\bibitem[\protect\citeauthoryear{B\"{o}hning, Diez, Scheub, Schlattmann, and
  Lindsay}{B\"{o}hning et~al.}{1994}]{BohDie94}
B\"{o}hning, D., E.~Diez, R.~Scheub, P.~Schlattmann, and B.~Lindsay (1994).
\newblock The distribution of the likelihood ratio for mixtures of densities
  from the one-parameter exponential family.
\newblock {\em Annals of the Institute of Statistical Mathematics\/}~{\em 46},
  373--388.

\bibitem[\protect\citeauthoryear{Bouveyron and Brunet-Saumard}{Bouveyron and
  Brunet-Saumard}{2014}]{Bouveyron13}
Bouveyron, C. and C.~Brunet-Saumard (2014).
\newblock Model-based clustering of high-dimensional data: A review.
\newblock {\em Computational Statistics and Data Analysis\/}~{\em 71}, 52--78.

\bibitem[\protect\citeauthoryear{Bouveyron, Girard, and Schmid}{Bouveyron
  et~al.}{2007}]{Bouveyron07}
Bouveyron, C., S.~Girard, and C.~Schmid (2007).
\newblock High-dimensional data clustering.
\newblock {\em Computational Statistics {\&} Data Analysis\/}~{\em 52\/}(1),
  502--519.

\bibitem[\protect\citeauthoryear{Browne and McNicholas}{Browne and
  McNicholas}{2015}]{BroMcN15}
Browne, R.~P. and P.~D. McNicholas (2015).
\newblock A mixture of generalized hyperbolic distributions.
\newblock  {\em Canadian Journal of Statistics\/}. In press.

\bibitem[\protect\citeauthoryear{Browne and McNicholas}{Browne and
  McNicholas}{2014}]{browne14}
Browne, R.~P. and P.~D. McNicholas (2014).
\newblock Estimating common principal components in high dimensions.
\newblock {\em Advances in Data Analysis and Classification\/}~{\em 8\/}(2),
  217--226.

\bibitem[\protect\citeauthoryear{Browne, McNicholas, and Sparling}{Browne
  et~al.}{2012}]{browne11}
Browne, R.~P., P.~D. McNicholas, and M.~D. Sparling (2012).
\newblock Model-based learning using a mixture of mixtures of {G}aussian and
  uniform distributions.
\newblock {\em IEEE Transactions on Pattern Analysis and Machine
  Intelligence\/}~{\em 34\/}(4), 814--817.

\bibitem[\protect\citeauthoryear{Campbell, Fraley, Murtagh, and
  Raftery}{Campbell et~al.}{1997}]{CamFra97}
Campbell, J.~G., F.~Fraley, F.~Murtagh, and A.~E. Raftery (1997).
\newblock Linear flaw detection in woven textiles using model-based clustering.
\newblock {\em Pattern Recognition Letters\/}~{\em 18\/}(1539--1548).

\bibitem[\protect\citeauthoryear{Chen, Cheung, So, Fan, Barry, Higgins, Lai,
  Ji, Dudoit, Ng, van~de Rijn, Botstein, and Brown}{Chen et~al.}{2002}]{chen02}
Chen, X., S.~T. Cheung, S.~So, S.~T. Fan, C.~Barry, J.~Higgins, K.-M. Lai,
  J.~Ji, S.~Dudoit, I.~O. Ng, M.~van~de Rijn, D.~Botstein, and P.~O. Brown
  (2002).
\newblock Gene expression patterns in human liver cancers.
\newblock {\em Molecular Biology of the Cell\/}~{\em 13\/}(6), 1929--1939.

\bibitem[\protect\citeauthoryear{Dasgupta and Raftery}{Dasgupta and
  Raftery}{1998}]{DasRaf98}
Dasgupta, A. and A.~E. Raftery (1998).
\newblock Detecting features in spatial point processed with clutter via
  model-based clustering.
\newblock {\em Journal of American Statistical Association\/}~{\em 93},
  294--302.

\bibitem[\protect\citeauthoryear{Dempster, Laird, and Rubin}{Dempster
  et~al.}{1977}]{DemLai77}
Dempster, A.~P., N.~M. Laird, and D.~B. Rubin (1977).
\newblock Maximum likelihood from incomplete data via the {EM} algorithm.
\newblock {\em Journal of the Royal Statistical Society: Series B\/}~{\em
  39\/}(1), 1--38.

\bibitem[\protect\citeauthoryear{Forina and Armanino}{Forina and
  Armanino}{1982}]{forina82b}
Forina, M. and C.~Armanino (1982).
\newblock Eigenvector projection and simplified non linear mapping of fatty
  acid content of Italian olive oils.
\newblock {\em Annali di Chimica\/}~{\em 72}, 127--141.

\bibitem[\protect\citeauthoryear{Forina, Armanino, Castino, and Ubigli}{Forina
  et~al.}{1986}]{ForArm86}
Forina, M., C.~Armanino, M.~Castino, and M.~Ubigli (1986).
\newblock Multivariate data analysis as a discriminating method of the origin
  of wines.
\newblock {\em Vitis\/}~{\em 25}, 189--201.

\bibitem[\protect\citeauthoryear{Forina and Tiscornia}{Forina and
  Tiscornia}{1982}]{forina82a}
Forina, M. and E.~Tiscornia (1982).
\newblock Pattern recognition methods in the prediction of {I}talian olive oil
  origin by their fatty acid content.
\newblock {\em Annali di Chimica\/}~{\em 72}, 143--155.

\bibitem[\protect\citeauthoryear{Franczak, McNicholas, Browne, and Murray}{Franczak
  et~al.}{2013}]{franczak13}
Franczak, B.~C., P.~D. McNicholas, R.~P. Browne, and P.~M. Murray (2013).
\newblock Parsimonious shifted asymmetric Laplace mixtures.
\newblock {arXiv} preprint: arXiv 1311.0317.

\bibitem[\protect\citeauthoryear{Franczak, Browne, and McNicholas}{Franczak
  et~al.}{2014}]{franczak14}
Franczak, B.~C., R.~P. Browne, and P.~D. McNicholas (2014).
\newblock Mixtures of shifted asymmetric {L}aplace distributions.
\newblock {\em IEEE Transactions on Pattern Analysis and Machine
  Intelligence\/}~{\em 36}(6), 1149--1157.

\bibitem[\protect\citeauthoryear{Ghahramani and Hinton}{Ghahramani and
  Hinton}{1997}]{ghahramani97}
Ghahramani, Z. and G.~E. Hinton (1997).
\newblock The {EM} algorithm for factor analyzers.
\newblock Technical Report CRG-TR-96-1, University of Toronto, Toronto.

\bibitem[\protect\citeauthoryear{Good}{Good}{1953}]{good53}
Good, I, J. (1953).
\newblock The population frequencies of species and the estimation of
  population parameters.
\newblock {\em Biometrika\/}~{\em 40}, 237--260.

\bibitem[\protect\citeauthoryear{Gorman and Sejnowski}{Gorman and
  Sejnowski}{1988}]{GorSej88}
Gorman, R.~P. and T.~J. Sejnowski (1988).
\newblock Analysis of hidden units in a layered network trained to classify
  sonar targets.
\newblock {\em Neural Networks, Vol. 1, pp.\/}~{\em 1}, 75--89.

\bibitem[\protect\citeauthoryear{Halgreen}{Halgreen}{1979}]{halgreen79}
Halgreen, C. (1979).
\newblock Self-decomposibility of the generalized inverse {G}aussian and
  hyperbolic distributions.
\newblock {\em Z.\ Wahrscheinlichkeitstheorie Verw.\ Gebiete\/}~{\em 47},
  13--18.
%

\bibitem[\protect\citeauthoryear{Hennig}{Hennig}{2010}]{henning2010}
Hennig, C. (2010).
\newblock Methods for merging Gaussian mixture components.
\newblock {\em Advances in Data Analysis and Classification\/}~{\em 4}, 3--34.

\bibitem[\protect\citeauthoryear{Hubert and Arabie}{Hubert and
  Arabie}{1985}]{HubAra85}
Hubert, L. and P.~Arabie (1985).
\newblock Comparing partitions.
\newblock {\em Journal of Classification\/}~{\em 2\/}(1), 193--218.

\bibitem[\protect\citeauthoryear{J{\o}rgensen}{J{\o}rgensen}{1982}]{jorgensen82}
J{\o}rgensen, B. (1982).
\newblock {\em Statistical Properties of the Generalized Inverse Gaussian
  Distribution}.
\newblock New York: Springer-Verlag.

\bibitem[\protect\citeauthoryear{Karlis and Santourian}{Karlis and
  Santourian}{2009}]{karlis09}
Karlis, D. and A.~Santourian (2009).
\newblock Model-based clustering with non-elliptically contoured distributions.
\newblock {\em Statistics and Computing\/}~{\em 19\/}(1),
  73--83.


\bibitem[\protect\citeauthoryear{Lee and McLachlan}{Lee and
  McLachlan}{2013a}]{lee13b}
Lee, S. and G.~McLachlan (2013a).
\newblock {\em EMMIXuskew: Fitting Unrestricted Multivariate Skew t Mixture
  Models}.
\newblock R package version~0.11-5.

\bibitem[\protect\citeauthoryear{Lee and McLachlan}{Lee and
  McLachlan}{2013b}]{lee13}
Lee, S.~X. and G.~J. McLachlan (2013b).
\newblock On mixtures of skew normal and skew t-distributions.
\newblock {\em Advances in Data Analysis and Classification\/}~{\em 7\/}(3),
  241--266.

\bibitem[\protect\citeauthoryear{Lin}{Lin}{2009}]{lin09}
Lin, T.-I. (2009).
\newblock Maximum likelihood estimation for multivariate skew normal mixture
  models.
\newblock {\em Journal of Multivariate Analysis\/}~{\em 100}, 257--265.

\bibitem[\protect\citeauthoryear{Lin}{Lin}{2010}]{lin10}
Lin, T.-I. (2010).
\newblock Robust mixture modeling using multivariate skew t distributions.
\newblock {\em Statistics and Computing\/}~{\em 20\/}(3), 343--356.

\bibitem[\protect\citeauthoryear{Lin, McLachlan, and Lee}{Lin
  et~al.}{2013}]{lin13}
Lin, T.-I., G.~J. McLachlan, and S.~X. Lee (2013).
\newblock Extending mixtures of factor models using the restricted multivariate
  skew-normal distribution.
\newblock {arXiv} preprint: arXiv 1307.1748.

\bibitem[\protect\citeauthoryear{Lin, McNicholas, and Hsiu}{Lin
  et~al.}{2014}]{lin14}
Lin, T.-I., P.~D. McNicholas, and J.~H. Hsiu (2014).
\newblock Capturing patterns via parsimonious t mixture models.
\newblock {\em Statistics and Probability Letters\/}~{\em 88}, 80--87.

\bibitem[\protect\citeauthoryear{Lindsay}{Lindsay}{1995}]{Lin95}
Lindsay, B. (1995).
\newblock Mixture models: Theory, geometry and applications.
\newblock In {\em NSF-CBMS Regional Conference Series in Probability and
  Statistics}, Volume~5, California: Institute of Mathematical Statistics:
  Hayward.

\bibitem[\protect\citeauthoryear{Lopes and West}{Lopes and
  West}{2004}]{LopWes04}
Lopes, H.~F. and M.~West (2004).
\newblock Bayesian model assessment in factor analysis.
\newblock {\em Statistica Sinica\/}~{\em 14}, 41--67.


\bibitem[\protect\citeauthoryear{Markos, {Iodice D'Enza}, and {Van de
  Velden}}{Markos et~al.}{2013}]{markos13}
Markos, A., A.~{Iodice D'Enza}, and M.~{Van de Velden} (2013).
\newblock {\em clustrd: Methods for joint dimension reduction and clustering}.
\newblock R package version 0.1.2.

\bibitem[\protect\citeauthoryear{Maugis, Celeux, and Martin-Magniette}{Maugis
  et~al.}{2009}]{maugis09}
Maugis, C., G.~Celeux, and M.~Martin-Magniette (2009).
\newblock Variable selection in model-based clustering: A general variable role
  modeling.
\newblock {\em Computational Statistics and Data Analysis\/}~{\em 53\/}(11),
  3872--3882.

\bibitem[\protect\citeauthoryear{McLachlan, Bean, and Jones}{McLachlan
  et~al.}{2007}]{mclachlan07}
McLachlan, G.~J., R.~W. Bean, and L.~B.-T. Jones (2007).
\newblock Extension of the mixture of factor analyzers model to incorporate the
  multivariate t-distribution.
\newblock {\em Computational Statistics and Data Analysis\/}~{\em 51\/}(11),
  5327--5338.


\bibitem[\protect\citeauthoryear{McLachlan and Peel}{McLachlan and
  Peel}{2000}]{mclachlan00a}
McLachlan, G.~J. and D.~Peel (2000).
\newblock Mixtures of factor analyzers.
\newblock In {\em Proceedings of the Seventh International Conference on
  Machine Learning}, San Francisco, pp.\  599--606. Morgan Kaufmann.

\bibitem[\protect\citeauthoryear{McLachlan, Peel, and Bean}{McLachlan
  et~al.}{2003}]{McLachlan03}
McLachlan, G.~J., D.~Peel, and R.~W. Bean (2003).
\newblock Modelling high-dimensional data by mixtures of factor analyzers.
\newblock {\em Computational Statistics and Data Analysis\/}~{\em 41},
  379--388.


\bibitem[\protect\citeauthoryear{McNicholas}{McNicholas}{2010}]{mcnicholas10b}
McNicholas, P.~D. (2010).
\newblock Model-based classification using latent {G}aussian mixture models.
\newblock {\em Journal of Statistical Planning and Inference\/}~{\em 140\/}(5),
  1175--1181.

\bibitem[\protect\citeauthoryear{McNicholas, Jampani, McDaid, Murphy, and
  Banks}{McNicholas et~al.}{2014}]{mcnicholas11}
McNicholas, P.~D., K.~R. Jampani, A.~F. McDaid, T.~B. Murphy, and L.~Banks
  (2014).
\newblock {\em {pgmm}: Parsimonious Gaussian Mixture Models}.
\newblock R package version~1.1.

\bibitem[\protect\citeauthoryear{McNicholas and Murphy}{McNicholas and
  Murphy}{2010}]{McNicholas10}
McNicholas, P.~D. and T.~B.~Murphy (2010).
\newblock Model-based clustering of microarray expression data via latent
  Gaussian mixture models.
\newblock {\em Bioinformatics\/}~{\em 26\/}(21), 2705--2712.

\bibitem[\protect\citeauthoryear{McNicholas and Murphy}{McNicholas and
  Murphy}{2008}]{McNicholas08}
McNicholas, P.~D.. and T.~B.~Murphy (2008).
\newblock Parsimonious {G}aussian mixture models.
\newblock {\em Statistics and Computing\/}~{\em 18\/}(3), 285--296.

\bibitem[\protect\citeauthoryear{McNicholas, McNicholas, and Browne}{McNicholas
  et~al.}{2013}]{smcnicholas13}
McNicholas, S.~M., P.~D. McNicholas, and R.~P. Browne (2013).
\newblock {Mixtures of variance-gamma distributions}.
\newblock Arxiv preprint arXiv:1309.2695.

\bibitem[\protect\citeauthoryear{Meng and {Van Dyk}}{Meng and {Van
  Dyk}}{1997}]{MenVan97}
Meng, X. and D.~{Van Dyk} (1997).
\newblock The {EM} algorithm-an old folk song sung to a fast new tune.
\newblock {\em Journal of the Royal Statistical Society: Series B (Statistical
  Methodology)\/}~{\em 59\/}(3), 511--567.

\bibitem[\protect\citeauthoryear{Montanari and Viroli}{Montanari and
  Viroli}{2011}]{Montanari11}
Montanari, A. and C.~Viroli (2011).
\newblock Maximum likelihood estimation of mixtures of factor analyzers.
\newblock {\em Computational Statistics and Data Analysis\/}~{\em 55},
  2712--2723.

\bibitem[\protect\citeauthoryear{Morris, McNicholas, and Scrucca}{Morris et~al.}{2013}]{morris13b}
Morris, K., P. D.~McNicholas, and L.~Scrucca (2013).
\newblock Dimension reduction for model-based clustering via mixtures of multivariate t-distributions.
\newblock {\em Advances in Data Analysis and Classification\/}~{\em 7}(3), 321--338.
  
\bibitem[\protect\citeauthoryear{Morris and McNicholas}{Morris and McNicholas}{2013}]{morris13a}
Morris, K. and P. D.~McNicholas (2013).
\newblock Dimension reduction for model-based clustering via mixtures of shifted asymmetric Laplace distributions.
\newblock {\em Statistics and Probability Letters\/}~{\em 83}(9),
  2088--2093.
  
\bibitem[\protect\citeauthoryear{Murray, Browne, and McNicholas}{Murray
  et~al.}{2013}]{murray13c}
Murray, P.~M., R.~B. Browne, and P.~D. McNicholas (2013).
\newblock Mixtures of `unrestricted' skew-t factor analyzers.
\newblock Arxiv preprint arXiv:1310.6224.

\bibitem[\protect\citeauthoryear{Murray, Browne, and McNicholas}{Murray
  et~al.}{2014a}]{murray14a}
Murray, P.~M., R.~B. Browne, and P.~D. McNicholas (2014a).
\newblock Mixtures of skew-t factor analyzers.
\newblock {\em Computational Statistics and Data Analysis\/}~{\em 77},
  326--335.

\bibitem[\protect\citeauthoryear{Murray, McNicholas, and Browne}{Murray
  et~al.}{2014b}]{murray14b}
Murray, P.~M., P.~D. McNicholas, and R.~B. Browne (2014b).
\newblock A mixture of common skew-$t$ factor analyzers.
\newblock {\em Stat\/}~{\em 3\/}(1), 68--82.

\bibitem[\protect\citeauthoryear{O'Hagan, Murphy, Gormley, McNicholas, and Karlis}{O'Hagan et al.}{2014}]{ohagan15}
OÕHagan, A., Murphy, T.~B., Gormley, I.~C., McNicholas, P.~D., and Karlis, D. (2014).
\newblock Clustering with the multivariate normal inverse Gaussian distribution.
\newblock {\em Computational Statistics and Data Analysis\/}. In press, \textsc{doi}: doi:10.1016/j.csda.2014.09.006 

\bibitem[\protect\citeauthoryear{R Core Team}{R Core Team}{2014}]{R14}
R Core Team (2014).
\newblock R: A Language and Environment for Statistical Computing.
\newblock R Foundation for Statistical Computing, Vienna, Austria.


\bibitem[\protect\citeauthoryear{Rand}{Rand}{1971}]{rand71}
Rand, W.~M. (1971).
\newblock Objective criteria for the evaluation of clustering methods.
\newblock {\em Journal of the American Statistical Association\/}~{\em 66},
  846--850.

\bibitem[\protect\citeauthoryear{Ritter}{Ritter}{2014}]{ritter14}
Ritter, G. (2014).
\newblock {\em Robust cluster analysis and variable selection}.
\newblock Chapman \& Hall/CRC Press, Boca Raton.

\bibitem[\protect\citeauthoryear{Rocci, Gattone, and Vichi}{Rocci
  et~al.}{2011}]{rocGat11}
Rocci, R., S.~A. Gattone, and M.~Vichi (2011).
\newblock A new dimension reduction method: Factor discriminant k-means.
\newblock {\em Journal of Classification\/}~{\em 28\/}(2), 210--226.

\bibitem[\protect\citeauthoryear{Schwarz}{Schwarz}{1978}]{Sch78}
Schwarz, G. (1978).
\newblock Estimating the dimension of a model.
\newblock {\em Annals of Statistics\/}~{\em 6}, 461--464.

\bibitem[\protect\citeauthoryear{Steane, McNicholas, and Yada}{Steane
  et~al.}{2012}]{steane12}
Steane, M.~A., P.~D. McNicholas, and R.~Yada (2012).
\newblock Model-based classification via mixtures of multivariate t-factor
  analyzers.
\newblock {\em Communications in Statistics -- Simulation and
  Computation\/}~{\em 41\/}(4), 510--523.

\bibitem[\protect\citeauthoryear{Subedi and McNicholas}{Subedi and
  McNicholas}{2014}]{subedi14}
Subedi, S. and P.~D. McNicholas (2014).
\newblock Variational {B}ayes approximations for clustering via mixtures of
  normal inverse {G}aussian distributions.
\newblock {\em Advances in Data Analysis and Classification\/}~{\em 8\/}(2),
  167--193.

\bibitem[\protect\citeauthoryear{Tan and Dowe}{Tan and Dowe}{2005}]{TanDow05}
Tan, P.~J. and D.~L. Dowe (2005).
\newblock {MML} inference of oblique decision trees.
\newblock {\em Advances in Artificial Intelligence\/}, 1082--1088.

\bibitem[\protect\citeauthoryear{Timmerman, Ceulemans, Roover, and
  Leeuwen}{Timmerman et~al.}{2013}]{TimCeu13}
Timmerman, M.~E., E.~Ceulemans, K.~Roover, and K.~Leeuwen (2013).
\newblock Subspace k-means clustering.
\newblock {\em Behavior Research Methods\/}, 1--13.

\bibitem[\protect\citeauthoryear{Tortora, Browne, Franczak, and
  McNicholas}{Tortora et~al.}{2015}]{tortora14}
Tortora, C., R.~P.~Browne, B.~C.~Franczak, and P.~D.~McNicholas (2015).
\newblock MixGHD: Model based clustering and classification using the mixture
  of generalized hyperbolic distributions.
\newblock {\em R package version~1.4\/}.

\bibitem[\protect\citeauthoryear{Vichi and Kiers}{Vichi and
  Kiers}{2001}]{Vickie01}
Vichi, M. and H.~Kiers (2001).
\newblock Factorial k-means analysis for two way data.
\newblock {\em Computational Statistics and Data Analysis\/}~{\em 37}, 29--64.

\bibitem[\protect\citeauthoryear{Vrbik and McNicholas}{Vrbik and
  McNicholas}{2012}]{vrbik12}
Vrbik, I. and P.~D. McNicholas (2012).
\newblock Analytic calculations for the {EM} algorithm for multivariate
  skew-mixture models.
\newblock {\em Statistics and Probability Letters\/}~{\em 82\/}(6), 1169--1174.

\bibitem[\protect\citeauthoryear{Vrbik and McNicholas}{Vrbik and
  McNicholas}{2014}]{vrbik14}
Vrbik, I. and P.~D. McNicholas (2014).
\newblock Parsimonious skew mixture models for model-based clustering and
  classification.
\newblock {\em Computational Statistics and Data Analysis\/}~{\em 71},
  196--210.

\bibitem[\protect\citeauthoryear{Wang, Ng  and McLachlan}{Wang et~al.}{2013}]{Wang13}
Wang, K., Ng, A., and G.~McLachlan (2013).
\newblock {\em EMMIXskew: The EM Algorithm and Skew Mixture Distribution}.
\newblock R package version~1.0.1.

\bibitem[\protect\citeauthoryear{Wei and McNicholas}{Wei and
  McNicholas}{2014}]{wei14}
Wei, Y. and P.~D. McNicholas (2014).
\newblock Mixture model averaging for clustering.
\newblock {\em Advances in Data Analysis and Classification\/}. To appear. \textsc{doi}: 10.1007/s11634-014-0182-6.

\bibitem[\protect\citeauthoryear{Woodbury}{Woodbury}{1950}]{woodbury50}
Woodbury, M. (1950).
\newblock Inverting modified matrices.
\newblock Technical Report~42, Princeton University, Princeton, N.J.

\bibitem[\protect\citeauthoryear{Zhou and Jiang}{Zhou and
  Jiang}{2004}]{ZhoJia04}
Zhou, Z.-H. and Y.~Jiang (2004).
\newblock Nec4. 5: neural ensemble based c4. 5.
\newblock {\em IEEE Transactions on Knowledge and Data Engineering\/}~{\em
  16\/}(6), 770--773.

\end{thebibliography}

\end{document}